\documentclass[10pt,english,final,a4paper]{article}
\usepackage[usenames,dvipsnames,svgnames]{xcolor}
\usepackage{hyperref} 
\hypersetup{
unicode=false,          
pdfauthor={JuanPi Carbajal},%
pdftitle={Mechanistic Emulator},%
colorlinks=true,       
linkcolor=OliveGreen,          
citecolor=Sepia,        
filecolor=magenta,      
urlcolor=NavyBlue,          
}

\usepackage[utf8]{inputenc}
\usepackage[T1]{fontenc}
\usepackage{textcomp}
\usepackage{lmodern} 
\usepackage{fouriernc} 

\usepackage{siunitx}
\usepackage[sort&compress]{natbib}

\usepackage{babel}
\usepackage{graphicx}
\usepackage[small]{caption}
\usepackage{bm}

\usepackage{amsmath}
\usepackage{amsfonts}
\usepackage{amssymb}
\usepackage{amsthm}
\usepackage{thmtools}
\usepackage[mathscr]{eucal}
\DeclareMathAlphabet{\mathcal}{OMS}{cmsy}{m}{n}

\usepackage{authblk}
\usepackage{multirow}

\usepackage{subcaption} 

\usepackage{enumitem}

\newcommand{\ud}{\mathrm{d}}

\newcommand{\tder}[2]{\frac{\ud{#1}}{\ud{#2}}}
\newcommand{\defitg}[4]{\int_{#1}^{#2} {#3}\ud {#4}}

\newcommand{\bmm}[1]{\bm{#1}}

\newcommand{\bset}[1]{\big\lbrace {#1} \big\rbrace}
\newcommand{\stimes}[2]{{#1}\!\times\!{#2}}
\newcommand{\trp}{\top}

\newcommand{\trset}[1]{{#1}_\text{\tiny trn}} 
\newcommand{\tsset}[1]{{#1}_\text{\tiny tst}} 

\DeclareMathOperator*{\sign}{sign}
\DeclareMathOperator*{\GP}{\mathcal{G}\!\mathcal{P}} 
\DeclareMathOperator*{\gaussian}{\mathcal{N}} 
\DeclareMathOperator*{\meanf}{\mathfrak{m}} 
\DeclareMathOperator*{\covf}{\mathfrak{K}} 
\DeclareMathOperator*{\Gf}{\bmm{G}} 
\DeclareMathOperator*{\Gfa}{\bmm{G}^{\dagger}} 

\newcommand{\szp}{m} 
\newcommand{\szP}{M} 
\newcommand{\nT}{N}  
\newcommand{\nD}{n}  

\newcommand{\pl}{\bm{\theta}} 
\newcommand{\pnl}{\bm{\gamma}} 
\newcommand{\plc}[1]{\theta_{#1}} 
\newcommand{\Pl}{\Theta} 
\newcommand{\Pnl}{\Gamma} 
\newcommand{\heavi}{\mathcal{H}}

\newcommand{\hl}[2]{\colorbox{#1}{#2}}
\newtheorem{problem}{Problem}

\defcitealias{Octave}{GNU Octave}
\defcitealias{Inkscape}{Inkscape}
\defcitealias{Sage}{Sage}

\begin{document}
\title{Appraisal of data-driven and mechanistic emulators of nonlinear hydrodynamic urban drainage
simulators}
\author[1]{Juan Pablo Carbajal}
\author[1]{João Paulo Leitão}
\author[1]{Carlo Albert}
\author[1]{Jörg Rieckermann}
\affil[1]{Swiss Federal Institute of Aquatic Science and Technology, Eawag, Überlandstrasse 133, 8600 Dübendorf, Switzerland}

\date{\today}
\maketitle

\abstract{
Many model based scientific and engineering methodologies, such as system identification, sensitivity analysis, optimization
and control, require a large number of model evaluations.
In particular, model based real-time control of urban water infrastructures and online flood alarm systems
require fast prediction of the network response at different actuation and/or parameter values.
General purpose urban drainage simulators are too slow for this application.
Fast surrogate models, so-called emulators, provide a solution to this efficiency demand.
Emulators are attractive, because they sacrifice unneeded accuracy in favor of speed.
However, they have to be fine-tuned to predict the system behavior satisfactorily.
Also, some emulators fail to extrapolate the system behavior beyond the training set.
Although, there are many strategies for developing emulators, up until now the selection of the emulation strategy remains subjective.
In this paper, we therefore compare the performance of two families of emulators for open channel flows in the context of urban drainage simulators.
We compare emulators that explicitly use knowledge of the simulator's equations, i.e. \emph{mechanistic emulators} based on Gaussian Processes, with purely data-driven emulators using \emph{matrix factorization}.
Our results suggest that in many urban applications, naive data-driven emulation outperforms mechanistic emulation.
Nevertheless, we discuss scenarios in which we think that mechanistic emulation might be favorable for i) extrapolation in time and ii) dealing with sparse and unevenly sampled data.
We also provide many references to advances in the field of Machine Learning that have not yet permeated into the Bayesian environmental science community.
}

\section{Introduction}

For many real-world systems with a nonlinear response, model based tasks such as sensitivity analysis, learning model parameters from data (i.e. system identification or model calibration), and real-time control, are hampered by the long runtime of the employed numerical simulators.
Even if runtimes are short, these methods require a large number of model runs, which can take a prohibitive long time.
One way of speeding up these tasks is to build fast surrogate models, so called \emph{emulators}, to replace the computationally expensive simulators.
An emulator is a numerical model that is tailored to approximate the results of a computationally expensive simulator with a huge reduction in the time needed to run a simulation~\citep{OHagan2006}, i.e. it is a metamodel.
These ideas also belong to the technique of Reduced-Order Models (ROM), specially for models based on Partial Differential Equations (PDE)~\citep{Baur2014, Quarteroni2016}, and emulation as described below.

\begin{figure}[htbp]
    \centering
    \includegraphics[width=0.8\columnwidth]{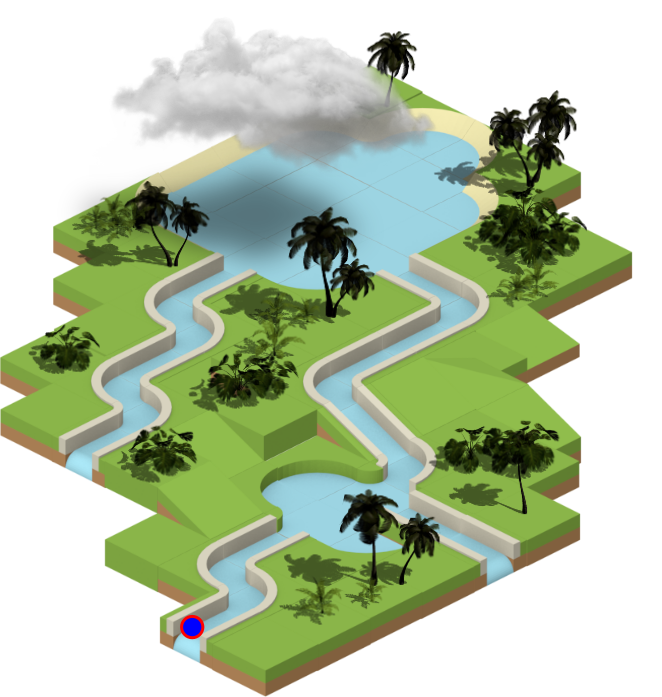}
    \protect\caption{\label{fig:gedankenE} An imaginary drainage network with flow limiting gate and/or water storage systems. The interesting singal is the level/flow at the outlet of the network, marked with a circle. Isometric tiles from \url{www.kenney.nl}.}
\end{figure}

To ground ideas, imagine that the flow at the outlet of a drainage network is limited using a flow limiting gate or by activating water storage systems (Fig.~\ref{fig:gedankenE}).
The position of the gate and the activation of storage is controlled using a model predictive controller~\citep{XI2013}.
Such a scenario is relevant in performance optimization of water treatment plants~\citep{fu2008multiobjective}.
The signals used to control the flows could be the current intensity and duration of rain events from several rain gauges within the catchment.
The controller needs to estimate an optimal course of action by predicting the flows induced by the rain and many possible actuations.
This optimization generally requires thousands of model runs, which can take a prohibitive long time when running a physically detailed simulator of the sewer network, such as a EPA Storm Water Management Model (SWMM) model~\citep{SWMM5}.
However, the simulator is just used to estimate the relation between the rain, the actuation, and the flow.
The full details of the simulator might not be required to obtain an accurate estimation of this relation.
Feedback control might further reduce the required accuracy of the estimated relation.

Emulation and interpolation are equivalent problems.
An emulator is built using the best available simulator to sample the space of actuations and/or parameters (henceforth the latter will include actuations).
The training data is then used to build an interpolation function which should predict values at unseen parameters with an acceptable degree of accuracy, which is case dependent.
That is, we reconstruct an unknown function $F:\mathbb{R}^{\vert\pnl\vert + 1} \rightarrow \mathbb{R}$, that takes a parameter vector of size $\vert\pnl\vert$ and a time instant, and generates the value of the magnitude of interest.
When this function is evaluated at the inputs used for training, the results are the same as the training outputs\footnote{Herein considered noiseless since they are generated by a deterministic simulator.}, i.e. this is the meaning of interpolation of the training data adopted herein.
Stated in this way, no distinction is needed between the parameters and the time components in the input.
However, knowing that the data is generated by a dynamical system, we separate time from the other parameter components.
Thus, we can find one interpolant in time and one in parameter space, which might be coupled to each other.
This parameter-time coupling emerges naturally in mechanistic emulation, as will be shown in Sec.~\ref{sec:mem}.

When the simulator is based on differential equations, the link between Gaussian Processes (GP) and linear stochastic differential equations (SDE)~\citep{Poggio1990, Steinke2008, Albert2012, Sarkka2013, Gonzalez2014, Solin2014}, permits the creation of GP based emulators that include knowledge about the simulator dynamics; these are called \emph{mechanistic emulators} (MEMs).
Conceptually, mechanistic emulation seeks a function that interpolates the training data whithin a class of functions defined by an SDE.
The importance of GPs for MEMs stems from the fact that they are the formal solution of this SDE.
Hence when the simulator is linear the emulator gives exact results; while for nonlinear simulators, the MEM will provide only an approximation.
Increasing the accuracy of this approximation and the efficiency of the methods are two fundamental challenges in GP based emulation~\citep[][Ch. 8]{Rasmussen06, OHagan2006}.

\citet{Reichert2011}~enumerated four overlapping approaches for developing emulators of dynamic simulators:

\begin{enumerate}[label=\roman*)]
\item \label{it:GP} Gaussian Processes
\item \label{it:BF} Basis function decompositions
\item State space transition function approximation
\item \label{it:KF} Stochastic linear model conditioned on data using Kalman smoothing.
\end{enumerate}
In particular approaches~\ref{it:GP} and~\ref{it:KF} are two different implementations of the same problem~\citep{Steinke2008}.
Roughly speaking the Kalman smoothing algorithm used in~\ref{it:KF} is an iterative implementation of the conditioning of the GP in~\ref{it:GP}.
The iteration in~\ref{it:KF} avoids the ill-conditioned covariance matrices~\citep{Hansen1998} involved in GP when sampling rates are high~\citep{Steinke2008,Reichert2011} and it is faster than direct matrix inversion in a serial implementation.
The GP approach~\ref{it:GP} is better suited for parallelization, speedups and energy saving via approximated computing~\citep{Angerer2015}.

Approaches~\ref{it:GP} (or~\ref{it:KF}) and~\ref{it:BF} are similar with respect to their implementation.
That is, approach~\ref{it:BF} can be implemented using GP regression~\citep[][sec. 2.7]{Rasmussen06}.
Therefore, the essential difference between~\ref{it:GP} and~\ref{it:BF} is that the former explicitly introduces mechanistic knowledge.
It will be shown here that approach~\ref{it:GP} is currently constrained to linear mechanistic knowledge, while popular methods based on maximum entropy~\citep{Victor1986,Christakos1998,Harte2011} can handle nonlinear knowledge.
The difference between GP based mechanistic emulation and maximum entropy methods is that in the latter, the mechanistic knowledge is added as constraints on the moments of the predictive or posterior distribution, while in the former its is added as the dynamics of the prior model.
Adding constraints to predictive distributions require expert knowledge available at the level of the emerging behavior of the simulator, while adding dynamic information requires knowledge about the constitutive elements parts of the simulator.
The latter is likely to be readily available from the development of the simulator itself.

Herein we compare the performance of GP emulators built using approaches~\ref{it:GP},  which we call \emph{mechanistic emulation}, and~\ref{it:BF} which we call \emph{data-driven emulation}.
The basis function that will be used for data-driven emulation will be derived solely from the data using matrix factorization, i.e. they will not explicitly include mechanistic knowledge.
In this article we use singular value decomposition (SVD) and nonnegative matrix factorization (NMF) to extract these bases (Sec.~\ref{sec:datadriven}), but more general basis extraction methods like Proper Orthogonal Decomposition (POD) could be used~\citep{Hesthaven2016}.
We compare results in an academic emulation problem to highlight the differences between the approaches (Sec.~\ref{sec:staticnl}-\ref{sec:staticnlC}).
We also provide emulation examples pertinent to the fields of hydrology and urban water management (Sec.~\ref{sec:wartegg}-\ref{sec:adliswil}).
In all of these, data-driven emulation outperforms mechanistic emulation.
The objective of this comparison is to provide intuition about the suitability of each approach, which is not available to date to the best of our knowledge, to highlight the need of enhancements of our emulators, and to motivate research questions in the field of emulation.
This is relevant, because, as described above, many applications in the field of urban drainage and flood predictions to date are hampered by slow models.
To a lesser degree, this is one of the few NMF applications in hydrology~\citep{Boian2014}.

\section{Methods and Materials}
\label{sec:methods}
In the subsequent sections we firstly describe the two emulation approaches used herein (Sec.~\ref{sec:datadriven}-\ref{sec:mem}).
Aiming at a wide readership, mathematical detail and rigor are kept at a minimum required level.
References are provided for the interested reader.
Secondly, we describe the datasets used for training and testing the emulators (Sec.~\ref{sec:dataset}).
These include models of two small catchments in Switzerland, that we use to thoroughly evaluate the performance of the emulators.

We use the word \emph{data} to refer to the input and output pairs provided by the simulator being emulated, e.g. in the case of a hydrodynamic simulator, inputs could be time and physical parameters of a sewer network, and outputs water levels or flows.

\subsection{Data-driven emulators}
\label{sec:datadriven}
In this approach we make the following assumptions about the simulation data used to build the emulator:

\begin{enumerate}[label=\alph*)]
\item The data contains the most significant dynamic features of the system response and these can be used as a \emph{time varying basis} to reproduce the data. \label{it:features}
\item Unseen system responses are well approximated by a linear combination of the features in~\ref{it:features}, i.e. there exist a solution manifold that can be well approximated with low dimensional sets~\citep{Quarteroni2016, Hesthaven2016}.
\item There exists a "smooth" mapping between inputs and the coefficients of the linear combinations of features.
\end{enumerate}

With these assumptions in mind we define the approximation strategy:

\begin{align}
y(t, \pnl) &\simeq \sum_{i=1}^{q} \beta_i(\pnl) \phi_i(t),\label{eq:decomp} \\
\beta_i(\pnl) &\sim \GP\left(\meanf{}_i(\pnl), \covf{}_i(\pnl,\pnl')\right)
\end{align}
\noindent where $y(t,\pnl)$ is the output of the simulator at time $t$ and parameters $\pnl$, $\beta_i$ is a mapping between these parameters and the components of the output in the basis function set ${\bset{\phi_i(t)}}$.
Following~\citet[][Sec. 2.7]{Rasmussen06} this could be generalized to a full GP regression problem.
However, as stated here the problem is simpler and it is justified by the performance it provides in the examples showcased in Sec.~\ref{sec:results}.

The procedure to build a data-driven emulator follows:

\begin{enumerate}[label=\roman*.]
\item \label{fit_step} Extract the first $q \leq \trset{\nD}$ features $\bset{\phi_i(t)}$ using the training set $\bset{y_i(t,\pnl_i)}_{i=1}^{\trset{\nD}}$.
This gives a $\stimes{q}{\trset{\nD}}$ matrix $\bmm{B}$ of coefficients ($q$ coefficients for each simulation used for training) and the basis evaluated at the observed time points $\Phi_{ij} = \phi_j(t_i)$, $i=1,\ldots,\nT$, $j=1,\ldots,q$.
\item \label{GP_step} Interpolate the $\bmm{B}$ coefficients from step~\ref{fit_step} using a GP to obtain a function of the inputs $\bmm{B} = f(\pnl)$.
\item Evaluate $f(\pnl)$ on the parameters in test set, use eq.~\eqref{eq:decomp} to predict outputs, and compare them with the output test set $\bset{y_i(t,\pnl_i)}_{i=1}^{\tsset{\nD}}$.
\end{enumerate}

The features $\bset{\phi_i(t)}$ are extracted from the training data itself via matrix factorization, which allows us to impose some general constraints on the features, e.g. nonnegativity.
Herein we use the singular value decomposition (SVD) and nonnegative matrix factorization (NMF), which are explained in later pragraphs.

These methods provide the basis evaluated only at the observed time points, $\Phi$, and to predict at unobserved times we linearly interpolate them over time.
The implications of this interpolation will be discussed in Sec.~\ref{sec:discussion}.
This approach decouples the interpolation in time with the interpolation in parameter space.

\paragraph{SVD}
is a robust factorization to calculate the eigenvectors of the covariance of the data matrix, i.e. the principal components.
Given the matrix $\bmm{Y} \in \mathbb{R}^{\stimes{\nT}{\nD}}$, SVD calculates the orthogonal matrices $\Phi \in \mathbb{R}^{\stimes{\nT}{\nD}}$ and $\bmm{W} \in \mathbb{R}^{\stimes{\nD}{\nD}}$, and the element-wise nonnegative diagonal matrix $\bmm{\Sigma} \in \mathbb{R}_{+}^{\stimes{\nD}{\nD}}$ with only $\Sigma_{ii} \neq 0$.
These matrices fulfill the relation
\begin{equation}
\bmm{Y} = \Phi \bmm{\Sigma} \bmm{W}^\trp.
\end{equation}
\noindent Using this factorization to calculate the covariance of the data matrix gives
\begin{equation}
\bmm{YY}^\trp = \Phi \bmm{\Sigma \Sigma}^\trp \Phi^\trp.
\end{equation}
\noindent Showing that $\Phi$ are eigenvectors of the covariance matrix, i.e the principal components of the data.
A decomposition of $\bmm{Y}$ in the form of eq.~\eqref{eq:decomp} is obtained by defining $\bmm{B}$ as the follows:
\begin{align}
\bmm{B} &= \bmm{\Sigma} \bmm{W}^\trp, \\
\bmm{Y} &= \Phi \bmm{B}.
\end{align}
\noindent The quality of approximation of the data degrades graciously with decreasing number of principal components.
Hence, only the first $q < \trset{\nD}$ principal components are used in general.
This reduces the size of the representation of the training data.
In the context of ROM and PDE, this decomposion is also know as Proper Orthogonal Decomposition (POD).

\paragraph{NMF}
provides an approximate minimum norm positive decomposition of the data~\citep{Kim2008}\footnote{function nmf\_bpas of GNU Octave's linear-algebra package \url{http://octave.sourceforge.net/linear-algebra/}.}.
Formally it solves the following problem:
\begin{problem}[NMF]
Given the matrix $\bmm{Y} \in \mathbb{R}^{\stimes{\nT}{\nD}}$ and $q \in \mathbb{N}$ with $q \leq \nD$, find matrices $\Phi \in \mathbb{R}_{+}^{\stimes{\nT}{q}}$ and $\bmm{B} \in \mathbb{R}_{+}^{\stimes{q}{\nD}}$ such that they minimize
\begin{equation}
\Vert \bmm{Y} - \Phi \bmm{B} \Vert^2 + a \Vert \Phi \Vert^2 + b \Vert \bmm{B} \Vert^2.
\end{equation}
\noindent Where $\Vert\cdot\Vert$ is the Frobenius matrix norm and $\mathbb{R}_{+}$ is the set of nonnegative real numbers.
\end{problem}

\noindent Intuitively, NMF works as SVD but constraining the principal components and the mixing coefficients to be nonnegative.
The decomposition controls the norm of the basis and its coefficients using two regularization terms parametrized with weights $a$ and $b$, which are manually tuned.
The nonnegative basis provided by NMF might be readily interpreted in physical terms in hydrological applications, yet the method is not widespread in the community.

\subsection{GP based mechanistic emulator (MEM)}
\label{sec:mem}
The predictive mean of a GP in $\bm{x}$ conditioned on data observed at $\bm{x}'$, has the following structure
\begin{align}
y(\bm{x}) &= \covf{(\bm{x},\bm{x}')}\bm{\alpha} + \meanf{(\bm{x})},\\
\bm{\alpha} &= \big(\covf{(\bm{x}',\bm{x}')} + \kappa I \big)^{-1} \big( y(\bm{x}') - \meanf{(\bm{x}')}\big).\label{eq:condition}
\end{align}
\noindent where $\covf$ and $\meanf$ stand for the covariance and the mean function of the prior GP, respectively.
The weights vector $\bm{\alpha}$ is learned from the data at the conditioning step, requiring the inversion of the matrix obtained by evaluating the covariance function at the observed inputs, shown in eq.~\eqref{eq:condition}.
The regularization parameter $\kappa$ encodes our trust on the prior knowledge and the error, if any, of the observations $y(\bm{x}')$\footnote{In the MEMs used herein, to obtain interpolation of the training data, we set the regularization parameter to the machine epsilon.}.

A MEM is the GP in time and simulator parameters associated with an input-driven linear stochastic ordinary differential equation (SODE) with a multidimensional state space.
Each component of the state space is defined by an observed simulator's parameter vector ($\pnl_i$).
Extra components are reserved for prediction at unseen simulator's parameter vectors\footnote{typically only one component is reserved for this, but more could be used in a parallel setting.}.

The process of building a MEM requires the definition of
\begin{enumerate}[label=\roman*.]
\item A linear prior model.
\item The covariance and mean functions of the GP.
\item A mapping from the parameters of the simulator to the parameters of the GP.
\end{enumerate}
\noindent In the following sections we obtain the mathematical expression of the covariance and mean functions of a first order linear time invariant (LTI) SODE, and explain the construction of the mapping between parameter spaces.
LTI systems often arise, for example, from a finite element modeling of partial differential equations.

For a rigorous and more general development, that include time varying parameters, see~\citet{Albert2012}.
The development for periodic difference equations was presented in~\citet{Steinke2008} and extended to a more general setting in~\citet{Gonzalez2014}.

A LTI SODE in the vector-valued function
\begin{equation}
\bm{s}(t): \mathbb{R}^+ \rightarrow \mathbb{R}^\szP,
\end{equation}

\noindent with initial condition $\bm{s}(0) = \bm{s}_0 \sim \mathcal{N}\left(\bm{\bar{s}}_0,\Sigma_0\right)$, is defined by the equation

\begin{equation}
\tder{\bm{s}}{t}(t) = \mathcal{A} \bm{s}(t) + \bm{u}(t) + \bm{\xi}(t),
\label{eq:sys1}
\end{equation}

\noindent where $\mathcal{A} \in \mathbb{R}^{\szP \times \szP}$, $\bm{u}(t)$ is a deterministic exogenous actuation, and $\bm{\xi}(t) \sim \mathcal{N}\left(0,\Sigma_\xi(t,t')\right)$ is a Gaussian noise term with covariance function
\begin{equation}
\Sigma_\xi(t,t') = \bmm{\Sigma} \delta(t-t') \quad \bmm{\Sigma} \in \mathbb{R}^{\szP \times \szP}.
\end{equation}

The general solution of this equation is
\begin{equation}
\bm{s}(t) = e^{\mathcal{A} t}\bm{s}_0 + \int_0^t e^{\mathcal{A} (t-\tau)} \left( \bm{u}(\tau) + \bm{\xi}(\tau) \right)\ud \tau.
\label{eq:sol1}
\end{equation}
\noindent where the first term is the solution of the homogenous ODE, i.e. with $\bm{u}(t)$ and $\bm{\xi}$ both zero.
The second term is the response of the ODE to the noisy actuation.

As mentioned above, this system depends on the simulator's parameters.
In general this means $\mathcal{A} (\pnl)$, $\Sigma_\xi(\pnl, \pnl')$, $\bm{u}(t, \pnl)$, and potentially $\bm{s}_0(\pnl)$.

\subsubsection{Covariance function}
\label{sec:covFunc}
We calculate the covariance function of the trajectories in eq.~\eqref{eq:sol1} using the formula
\begin{equation}
\covf{(t,r)} = \mathbb{E}\left[\left(\bm{s}(t)-\mathbb{E}[\bm{s}(t)]\right)\left(\bm{s}(r)-\mathbb{E}[\bm{s}(r)]\right)^\trp\right],
\label{eq:covFuncdef}
\end{equation}
\noindent where $\bm{s}^\trp$ indicates transposition and $\mathbb{E}$ is the ensemble average over initial conditions and noise realizations.

The ensemble average of the solutions, $\mathbb{E}[\bm{s}(t)]$, is solely determined by the average of the homogeneous part (depending only on the initial condition) and the deterministic actuation $\bm{u}$, because the contribution of the noise term $\bm{\xi}$ vanishes due to its zero mean value:
\begin{equation}
\mathbb{E}[\bm{s}(t)] = e^{\mathcal{A} t} \bm{\bar{s}}_0 + \int_0^t e^{\mathcal{A} (t-\tau)}\bm{u}(\tau)\ud \tau.
\label{eq:avg1}
\end{equation}
\noindent That is, the mean function is the solution to the deterministic (noise-free) inhomogeneous ODE.

From eqs.~\eqref{eq:sol1} and~\eqref{eq:avg1} we obtain the factors in the expectation in eq.~\eqref{eq:covFuncdef},
\begin{equation}
\bm{s}(t)-\mathbb{E}[\bm{s}(t)] = e^{\mathcal{A} t}\left(\bm{s}-\bm{\bar{s}}_0\right) + \int_0^t e^{\mathcal{A} (t-\tau)}\bm{\xi}(\tau)\ud \tau.
\end{equation}
\noindent Inserting this in the expression for the covariance function, eq.~\eqref{eq:covFuncdef},  we obtain four terms.
The first term is the covariance of the initial conditions.
The second term is a product of the integral of the noise term.
The last two terms are products of the initial conditions and the noise term; these terms will vanish due to the independence of the initial condition and the noise term, and due to the zero mean of the latter.
Finally, we obtain:

\begin{equation}
\begin{split}
\covf{(t,r)} = e^{\mathcal{A}t} \Sigma_0 e^{\mathcal{A}^\trp r} + \\
\int_0^t\int_0^r & e^{\mathcal{A}(t-\tau)} \Sigma \delta(\tau-\rho) e^{\mathcal{A}^\trp(r-\rho)} \ud \rho \ud \tau = \\
 e^{\mathcal{A}t} \Sigma_0 e^{\mathcal{A}^\trp r} + &\int_0^{\min(t,r)} e^{\mathcal{A} (t-\mu)} \Sigma e^{\mathcal{A}^\trp (r-\mu)} \ud \mu.
\end{split}
\label{eq:covFunc}
\end{equation}

The second term can be recognized as the property of the covariance under a linear
transformation:
\begin{equation}
\bm{s} = \Gf \bm{v} \quad \bmm{\Sigma_s} = \Gf \left(\Gf \bmm{\Sigma_v}\right)^\trp,
\label{eq:covprop}
\end{equation}
\noindent where the matrix product should be interpreted as the application of the integral
\begin{align}
(\bmm{A} \bm{v})(t) = \defitg{0}{\infty}{\bmm{A}(t,\mu) \bm{v}(\mu)}{\mu}, \\
(\bmm{A} \bmm{B})(t,r) = \defitg{0}{\infty}{\bmm{A}(t,\mu) \bmm{B}(\mu, r)}{\mu}.
\end{align}

In our case $\Gf$ stands for the Green's function of the linear operator
\begin{equation}
\tder{}{t} - \mathcal{A},
\end{equation}
\noindent which is
\begin{equation}
\Gf(t,r) = \heavi(t-r)e^{\mathcal{A} \left(t - r\right)},
\end{equation}
\noindent with $\heavi$ the Heaviside or Step function. Its transpose (adjoint) is
\begin{equation}
\Gfa(t,r) = \heavi(r-t)e^{\mathcal{A}^{\trp} \left(r - t\right)}.
\end{equation}

\noindent giving
\begin{equation}
\begin{split}
&\left(\bmm{G} \left(\bmm{G} \bmm{\Sigma_\xi} \right)^{\dagger}\right)(t,r) = \\
&\defitg{0}{\infty}{\Gf(t,\mu) \left(\defitg{0}{\infty}{\Gf(\mu,\tau)\bmm{\Sigma} \delta(\tau - r)}{\tau}\right)^{\dagger}}{\mu} = \\
&\defitg{0}{\infty}{\Gf(t,\mu) \bmm{\Sigma}\Gfa(\mu,r)}{\mu} = \\
&\defitg{0}{\infty}{\heavi(t-\mu)\heavi(r-\mu) e^{\mathcal{A} \left(t - \mu\right)} \bmm{\Sigma} e^{\mathcal{A}^\trp \left(r - \mu\right)}}{\mu} =\\
&\int_0^{\min(t,r)} e^{\mathcal{A} (t-\mu)} \bmm{\Sigma} e^{\mathcal{A}^\trp (r-\mu)} \ud \mu
\end{split}
\end{equation}

This implies that if the differential operator has a known Green's function,
then the covariance function of the GP can be calculated by transforming the covariance function $\Sigma$ of the noisy input $\bm{\xi}$.
A more general and formal treatment of this process is described in~\citet{Kimeldorf1970} and the relation between differential operators and kernels is summarized in~\citet{Steinke2008}.
This covariance function computes the statistical interactions between the components of the LTI SODE.
In other words, it provides the coupling of components of what is know as multi-output GP in the machine learning community and cokriging in geostatistics~\citep[][Sec. 9.1]{Rasmussen06}.
In~\citet{Fricker2013} different structures of the coupling between output components were studied, MEMs automatically derive the coupling structure from the available prior knowledge.

\subsubsection{Linear prior}
To determine the elements in the system matrix $\mathcal{A}$ in~\eqref{eq:sys1}, we select a linear model for each output in the training data corresponding to a simulator's parameter vector, which we call the \emph{linear proxy}:
\begin{align}
\mathcal{L}_i: \mathbb{R}_+ \times \mathbb{R}^q &\rightarrow \mathbb{R}^d, \\
t, \pl_i &\mapsto \bm{z}(t, \pl_i).
\end{align}
\noindent The $q$ dimensional parameter vector $\pl_i$ of the proxy depends on the simulator's parameters used for the $i$th simulation, i.e. $\pl_i(\pnl_i)$.
The proxy's output dimension $d$ is equal to dimension of each output in the training data, usually $d=1$.
The concatenated set of proxies defines the \emph{linear prior} of the MEM.

Herein, the proxies will be $\szp$ dimensional linear time invariant (LTI) ODEs, i.e. in state space form
\begin{align}
\tder{\bm{s}}{t}(t) &= \bmm{A}(\pl) \bm{s}(t) + \bm{u}(t, \pl) \quad \bmm{A} \in \mathbb{R}^{\szp\times \szp}, \bm{u} \in \mathbb{R}^{\szp}\\
\bm{z}(t) &= \bmm{C} \bm{s}(t)\phantom{(\pl)+\bm{u}(t, \pl)} \quad \bmm{C} \in \mathbb{R}^{d\times \szp}
\end{align}
\noindent Here all proxies have the same dimension $\szp$.
Although this is not required by the method, it is the simplest structure of MEMs~\citep{Albert2012}.
Nevertheless, proxies with different dimensions might be better suited for dynamical systems with bifurcations, e.g. training data containing a mixture of oscillating and converging time series, due to the presence of a Supercritical Andronov-Hopf bifurcation~\citep{Kuznetsov2006} in the simulator.

The emulator's linear prior is constructed by aggregating the proxies and by coupling them with a noise term with a covariance that depends on the simulator's parameters, i.e
\begin{align}
\tder{S}{t}(t) &= \mathcal{A}(\Pl) S(t) + U(t, \Pl) + \bm{\xi}(t, \Pnl),\\
Z(t) &= \mathcal{C}(\Pl) S(t),\\
\mathcal{A} &\in \mathbb{R}^{\stimes{\szp\nD}{\szp\nD}}, \; U,\xi \in \mathbb{R}^{\stimes{\szp\nD}{1}}\, \mathcal{C} \in \mathbb{R}^{\stimes{d}{\szp\nD}}.\nonumber
\end{align}
\noindent Where $\Pl = \left\lbrace\pl_i\right\rbrace$ and $\Pnl = \left\lbrace\pnl_i\right\rbrace$ contain the corresponding $\nD$ parameter vectors. The matrix $\mathcal{A}$ is block diagonal, $\mathcal{C}$ is a horizontal concatenation, and $U$ a vertical concatenation:

\begin{align}
\begin{split}
\mathcal{A}(\Pl) &= \begin{bmatrix} \bmm{A}(\pl_1) & &\\ & \ddots & \\ & & \bmm{A}(\pl_\nD)\end{bmatrix},\\
U(t,\Pl) &= \begin{bmatrix} \bm{u}(t,\pl_1) \\ \vdots \\ \bm{u}(t, \pl_\nD)\end{bmatrix},\\
\mathcal{C}(\Pl) &= \begin{bmatrix} \bmm{C}(\pl_1) & \hdots & \bmm{C}(\pl_\nD)\end{bmatrix}.
\end{split}\label{eq:matrices}
\end{align}

\subsubsection{Parameter mapping}
To evaluate the matrices in eq.~\eqref{eq:matrices} we need a mapping from simulator's parameters $\pnl$ to proxy's parameters $\pl$, , i.e. $\Pnl \mapsto \Pl$.
It can be an ad-hoc function derived from knowledge about the simulator, as was done in~\citep{Machac2016, Machac2016b, Albert2012} or it can be learned directly from the data.
The latter is especially useful if proxies with different dimensions are combined to form the linear prior of the emulator.

A proxy's parameter $\pl_i$ can be learned from the data via optimization, e.g. least squares fit of the data $y_i(T, \pnl_i)$ using $\bm{z}(T,\pl_i)$.
Alternatively, $\Pl$ can be left as hyperparameters in the GP and estimated by maximizing the likelihood.
In any case, the obtained pairs $(\pnl_i, \pl_i)$ are then used to learn a mapping between the simulator's and emulator's parameter spaces.

\subsubsection{Warped MEM}\label{sec:WGP}
It is possible to add some nonlinear knowledge to a mechanistic emulator when the simulator's equations can be approximated with a Wiener model, i.e. a time independent nonlinear function applied to the states of a linear dynamical system.
To do this we use Warped GPs~\citep{Snelson03}.
This modification does not affect the structure of the emulator, only the data on which it is trained.
In addition to the linear dynamical system parameters defining the prior, the parameters of the nonlinear function need to be learned as well unless this mapping is explicitly given, e.g. from the linearization of a nonlinear ODE via a change of variables as in the Bernoulli ODE~\citep{Hairer1993}.

\subsection{Datasets description}
\label{sec:dataset}
Simulated data used to build emulators are provided in a set of scalar signals $y_i(t_j, \pnl_i)$ with $i=1,\ldots,\nD$, all of them sampled at the same time points, i.e. $T = \bset{t_j},\; j=1,\ldots,\nT$.
For all emulators reported here, the time series in the datasets were subsampled as much as possible, without deteriorating their relevant dynamic features, e.g. oscillations and/or peaks.
In all cases, the datasets are randomly separated in a training set of size $\trset{\nD}$ and a test set of size $\tsset{\nD}$, with $\trset{\nD} + \tsset{\nD} = \nD$.
Table~\ref{tab:dataset} summarizes the properties of the data used.

\vspace{1em}
\begin{table*}[th]
\centering
\begin{tabular}{lccc}
Dataset & Nonlinear DS I \& II & Wartegg & Adliswil\\
\hline
\# parameters, $\vert\pnl\vert$ & 1 & 2 & 8\\
Time samples, $\nT$(used/total) & 6/40 & 52/2880 & 193/601 \\
Training set, $\trset{\nD}$ & 10 & 200 & 128\\
Test set, $\tsset{\nD}$ & 190 & 700 & 128\\
\end{tabular}
\caption{\label{tab:dataset} Summary of datasets used herein.}
\end{table*}

\subsubsection{Nonlinear dynamical system dataset I \& II}
\label{sec:staticnl_method}
To illustrate the virtues of MEMs, we first consider a didactical example using data generated by the model

\begin{align}
\tder{x}{t} &= a(x_0) x + b(x_0), \quad x(0) = x_0, \label{eq:staticnl1}\\
a(x) &= - a_0 e^{- a_1 \left(x - \sign{(x)}\right)^2},\label{eq:staticnl2}\\
b(x) &= b_0 \tanh{(x)}.\label{eq:staticnl3}
\end{align}

\noindent The parameters values are $a_0=12.616, a_1=5, b_0=2$.

The time evolution of this contrived system is linear.
The parameters defining the evolution depend nonlinearly on the initial condition, i.e. the parameters of the linear system are nonlinear functions of the initial condition.
Therefore an emulator of this system takes time and the initial condition ($\vert\pnl\vert=1$) as inputs.
The behavior of this system meets the above assumptions underlying a MEM with a linear time-invariant prior.
Thus an emulator without coupling noise solves this system exactly.

The first dataset consists of simulated time series with 40 output observations for \num{200} initial conditions $x_0 \in [-1,1]$.

The second dataset is built by mixing the trajectories of the dynamical system in eq.~\eqref{eq:staticnl1}, specifically:
\begin{align}
\hat{x}(t,x_0) = \int_{\alpha} x(t,x_0+\alpha) \gaussian_\alpha(x_0,\sigma^2) \ud \alpha\label{eq:staticnlC},
\end{align}
\noindent where $\sigma = 0.5$.
Although the nonlinearities in the dynamical system still depend only on the initial condition, the smoothing couples neighboring trajectories.

For the mechanistic emulation we will use a $1$ dimensional linear proxy:
\begin{equation}
\tder{s}{t}(t) = \plc{1}(s_0) s(t) + \plc{2}(s_0),
\end{equation}
\noindent and use the knowledge from the first simulator to set $s_0 = x_0$, $\plc{1}(s_0) = a(s_0)$, and $\plc{2}(s_0) = b(s_0)$.
Since each proxy is the solution of the system in eq.~\eqref{eq:staticnl1}, no coupling of the components of the MEM is needed in the first case, i.e. $\bmm{\Sigma} = \bmm{I}$.
For the second simulator we will couple the components using a $1$ dimensional Matérn covariance function and the parameters mapping will be learned from the data.

\subsubsection{Wartegg catchment dataset}
\label{sec:wartegg_m}
This dataset was generated from a SWMM model of a \SI{2.64}{\kilo\metre\squared} urban catchment located in the city of Lucerne in the canton of Lucerne, Switzerland.
This model has been calibrated satisfactorily to observed rainfall-runoff and used for hydrological studies of the site~\citep[][detailed model description provided therein]{Tokar2015}.

The dataset consists of \num{900} time series with \num{2880} data points, simulating \SI{24}{\hour} of water levels in an open outlet during different rain events.
To drive the system into a highly nonlinear behavior we synthetically generated rain events covering a wide range of return periods.
These events were generated following a block rain model~\citep[][Ch. 13.2]{Gjr2007} parametrized by intensity ($I$) and duration ($d$), i.e. $\pnl = (I,d)$ and $\vert\pnl\vert=2$.
The intensity of the event spans \num{30} values in the range $I \in \SIrange{10}{100}{\milli\metre\per\hour}$, with \num{30} different durations ($d \in \SIrange{10}{240}{\minute}$) for each intensity.

For the mechanistic emulation we will use a $1$ dimensional linear proxy for the water level:
\begin{equation}
\tder{h}{t}(t) = \plc{1}(\pnl) h(t) + \plc{2}(\pnl) R(t, \pnl)\\
\end{equation}
\noindent where $R(t, \pnl)$ is the rain event used in the simulation.
The values of $\plc{1}(\pnl)$ and $\plc{2}(\pnl)$ are obtained by a least squares fit of the data.

Due to the low performance achieved by standard MEMs in this dataset, the actual MEM presented in Sec.~\ref{sec:wartegg} uses a Wiener model as proxy.
This is done using Warped GPs (Sec.~\ref{sec:WGP}), with a nonlinearity given by $\hat{h}(t) = a(\pnl)\tanh(b(\pnl) h(t) + c(\pnl))$, where the parameters $a$, $b$ and $c$ are learned from the data.

\subsubsection{Adliswil catchment dataset}
This dataset was generated from a SWMM model of a \SI{1.63}{\kilo\metre\squared} urban catchment located in the city of Adliswil in the canton of Zurich, Switzerland.
The dataset was used in~\citet{Machac2016}(detailed model description provided therein) to create an emulator which was then used to speed-up the calibration (identification) of the parameters in the simulator.

The dataset consist of \num{256} time series with \num{601} data points, all of them corresponding to a single rain event, but with different $\num{8}$ dimensional input parameter vectors ($\vert\pnl\vert=\num{8}$).
The time series are simulations of inflow to the local WWTP, and the parameters describe the physical properties of the sewer network.

For the mechanistic emulation we will use the same linear proxy as in~\citet{Machac2016}, a $1$ dimensional ODE for the discharge:
\begin{equation}
\tder{Q}{t}(t) = \plc{1}(\pnl) Q(t) + \plc{2}(\pnl) R(t)\\
\end{equation}
\noindent where $R(t)$ is the measured rain event.

Two MEMs will be built, the first uses the values of $\plc{1}(\pnl)$ and $\plc{2}(\pnl)$ using the relation from~\citet{Machac2016}, and the second will obtain them from a least squares fit of the data.

\subsection{Performance assessment}

To asses the quality of an emulator we compute the following emulation errors on the data reserved for testing:
\begin{enumerate}[label=\roman*.]
\item Maximum absolute error (MAE)\label{MAE}
\begin{equation}
e_{\scriptsize{\text{MAE}}}(\pnl_j) = \max_i \left\vert \hat{y}(t_i,\pnl_j) - y(t_i,\pnl_j) \right\vert
\end{equation}
\noindent where $\hat{y}(t,\pnl)$ and $y(t,\pnl)$ are the emulated and simulated responses, respectively.
\item Root mean square error (RMSE)\label{RMSE}
\begin{equation}
e_{\scriptsize{\text{RMSE}}}(\pnl_j) = \sqrt{ \frac{1}{\nT}\sum_{i=1}^{\nT} \left(\hat{y}(t_i,\pnl_j) - y(t_i,\pnl_j)\right)^2}
\end{equation}
\end{enumerate}
Error $e_{\scriptsize{\text{MAE}}}$ measures the error in reproducing extreme values and/or peaks in the simulated signals, while error $e_{\scriptsize{\text{RMSE}}}$ measures an overall quality of the emulation.

\section{Results}
\label{sec:results}

\subsection{Nonlinear system dataset}
\label{sec:staticnl}
This dataset is used to highlight the value of the mechanistic over data-driven emulation.
It is suited for illustration purposes, since the surface to be reconstructed can be plotted ($\vert\pnl\vert=1$).

\begin{figure}[htbp]
    \centering
    \begin{subfigure}[b]{\columnwidth}
        \includegraphics[width=\columnwidth]{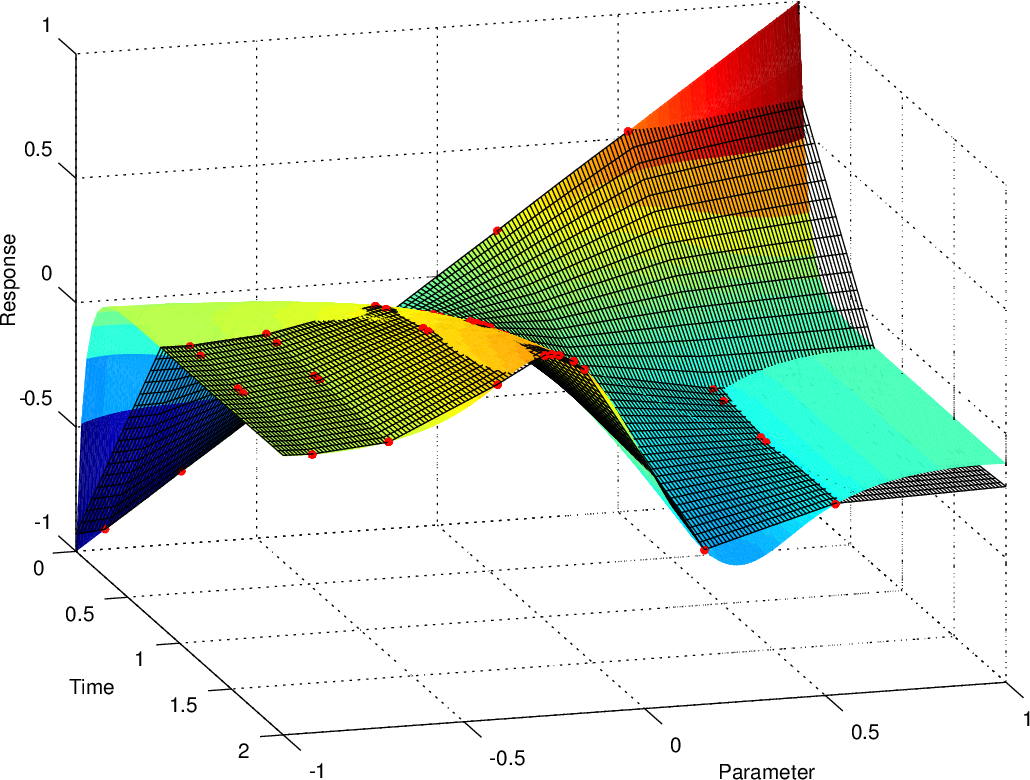}
        \caption{SVD}
        \label{fig:staticnl_svd}
    \end{subfigure}
    \\
    \begin{subfigure}[b]{\columnwidth}
        \includegraphics[width=\columnwidth]{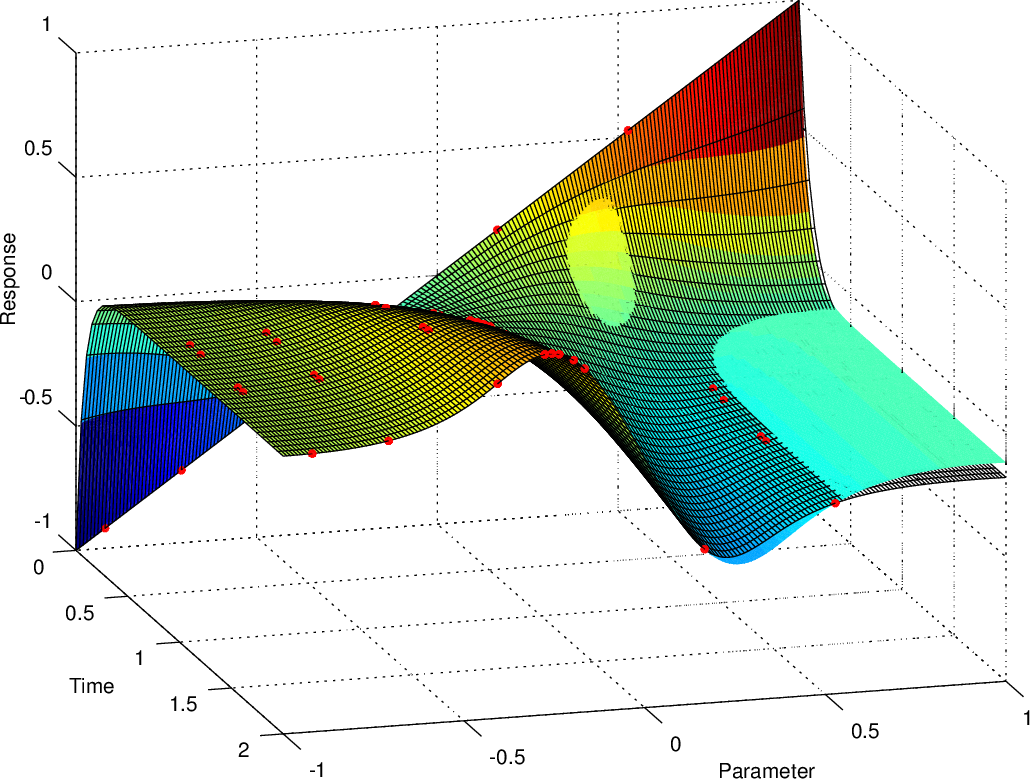}
        \caption{MEM}
        \label{fig:staticnl_mem}
    \end{subfigure}
    \protect\caption{\label{fig:staticnl}Nonlinear system emulation of didactical model.
    The colored surface shows the behaviour of the simulator output.
    Red dots show the training data.
    The emulated surface is shown with a black wireframe.
    In this contrived scenario a MEM~(\subref{fig:staticnl_mem}) outperforms an SVD emulator~(\subref{fig:staticnl_svd}).
    The failure of SVD is mainly due to the linear interpolation in the time direction.}
\end{figure}

Fig.~\ref{fig:staticnl} shows the response of the system described in eqs.~\eqref{eq:staticnl1}-\eqref{eq:staticnl3} as a surface $\mathbb{R}^2\rightarrow\mathbb{R}$.
Fig.~\ref{fig:staticnl_svd} illustrates the weakness of the SVD emulator, which does not exploit the simulator's parameter dependence.
Moreover predictions at unseen time points are provided by linear interpolation of the SVD basis.
To build the MEM we used the exact simulator's parameter dependence.
Although training data is sparse in the time direction, since the MEM encodes the right time evolution, the reconstruction quality is high, as can be seen in Fig.~\ref{fig:staticnl_mem}.
Although the SVD emulator could be improved by using an exponential basis for time interpolation~\citep{franzchoose}, i.e. the one provided by the covariance function of the MEM, the recovered parameter dependence will still be poor due to the low sampling of the parameter space.

This example shows that mechanistic emulation is ideal for situations where there is good prior knowledge and the training data is sparse.
This is even more striking when data corresponding to different simulator parameters is sampled at different times, e.g. with adaptive stepsize simulators.
In this case mechanistic emulation can be applied directly, while SVD emulation becomes complicated as a matrix completion problem needs to be solved before factorization can be applied~\citep[see][for an SVD relevant analysis]{oh2010matrix}.
In a similar fashion the Kalman smoothing algorithm described in~\citet{Reichert2011} also requires a first step in which all the data is interpolated into the same temporal grid, e.g. via interpolation.

\subsection{Nonlinear system dataset II}
\label{sec:staticnlC}

\begin{figure}[htbp]
    \centering
    \begin{subfigure}[b]{\columnwidth}
        \includegraphics[width=\columnwidth]{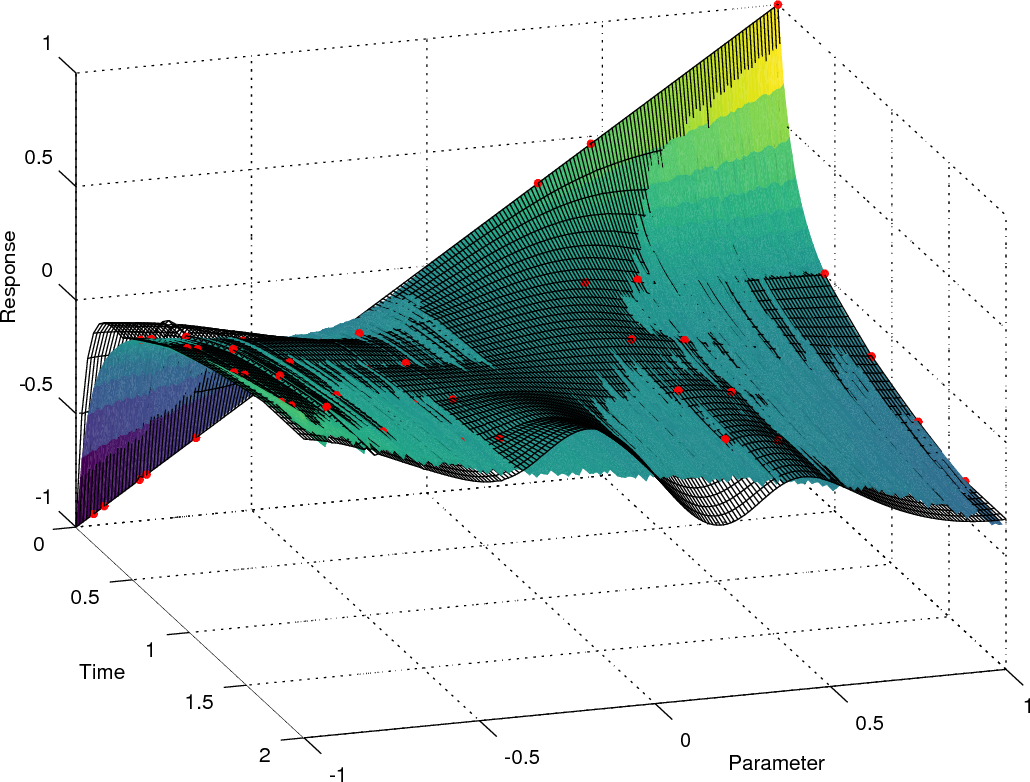}
        \caption{MEM}
        \label{fig:staticnlC_mem}
    \end{subfigure}
    \\
    \begin{subfigure}[b]{\columnwidth}
        \includegraphics[width=\columnwidth]{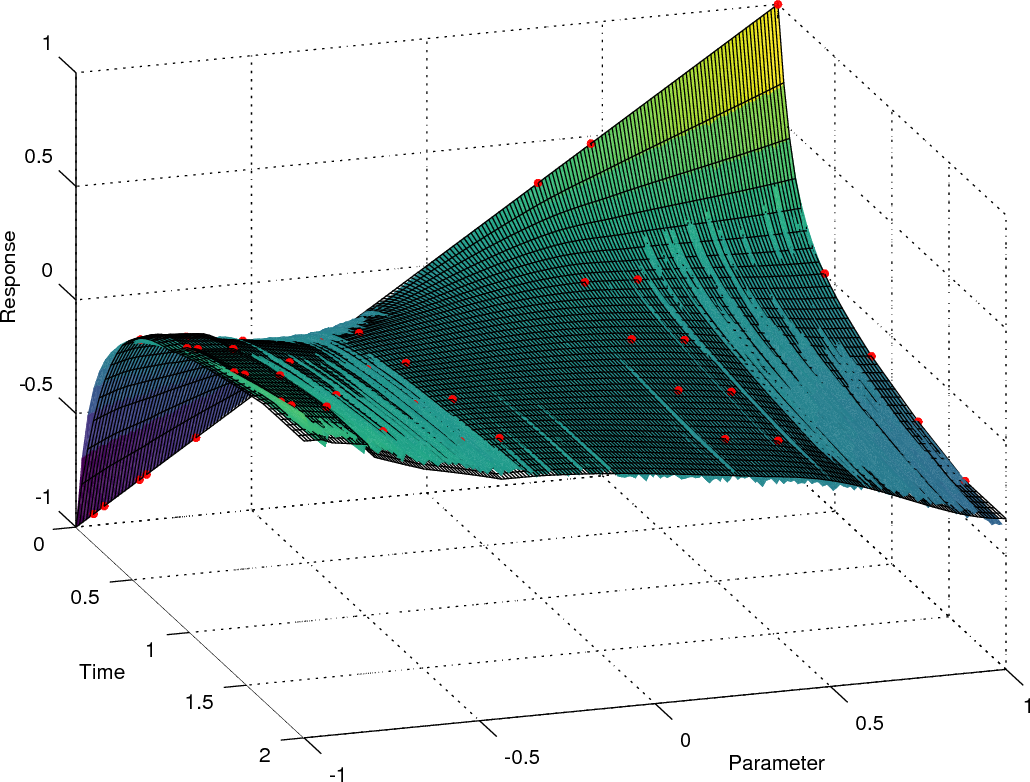}
        \caption{MEM fitted proxy}
        \label{fig:staticnlC_memf}
    \end{subfigure}

    \protect\caption{\label{fig:staticnlC}Nonlinear system emulation II.
    The colored surface shows the surface to reconstruct.
    In this second contrived scenario the MEM~(\subref{fig:staticnl_mem}) performs the same as the SVD emulator~(\subref{fig:staticnl_svd}).
    Red dots show the training data.
    The emulated surface is shown with a black wireframe.
    }
\end{figure}

Since the structure of the proxy of a MEM does not change once its dimension $\szp$ is set, we can optimize the proxy's parameters to reduce epistemic biases.
The example shown in Fig.~\ref{fig:staticnlC} illustrates the effect of mismatches between the dynamical system and the proxy, i.e. epistemic bias, and how it can be mitigated.
In Fig.~\ref{fig:staticnlC_mem} we show the performance of a MEM built with the same proxy as in Fig.~\ref{fig:staticnl_mem}, but used to emulate the nonlinear dynamical system described at the end of Sec.~\ref{sec:staticnl_method}.
Comparing with Fig.~\ref{fig:staticnl_mem}, we see that in the regions where there is no observed data, the emulation is biased.
A MEM with proxies fitted to the data is shown in Fig.~\ref{fig:staticnlC_memf}.
In this case the epistemic bias is considerably reduced although we did not used mechanistic knowledge to improve the parameters mapping.

Hence, if the mapping between the simulator parameters and the linear proxy's parameters cannot be exactly determined from the mechanistic knowledge, it is better to fit the proxy's parameters to the data instead of using and ad-hoc calculation based on on one's best guess or expert opinion.
The latter can be improved a posteriori by studying the parameters mapping emerging from the fit.

\subsection{Wartegg catchment dataset}
\label{sec:wartegg}

\begin{figure}[htbp]
\centering
\includegraphics[width=\columnwidth]{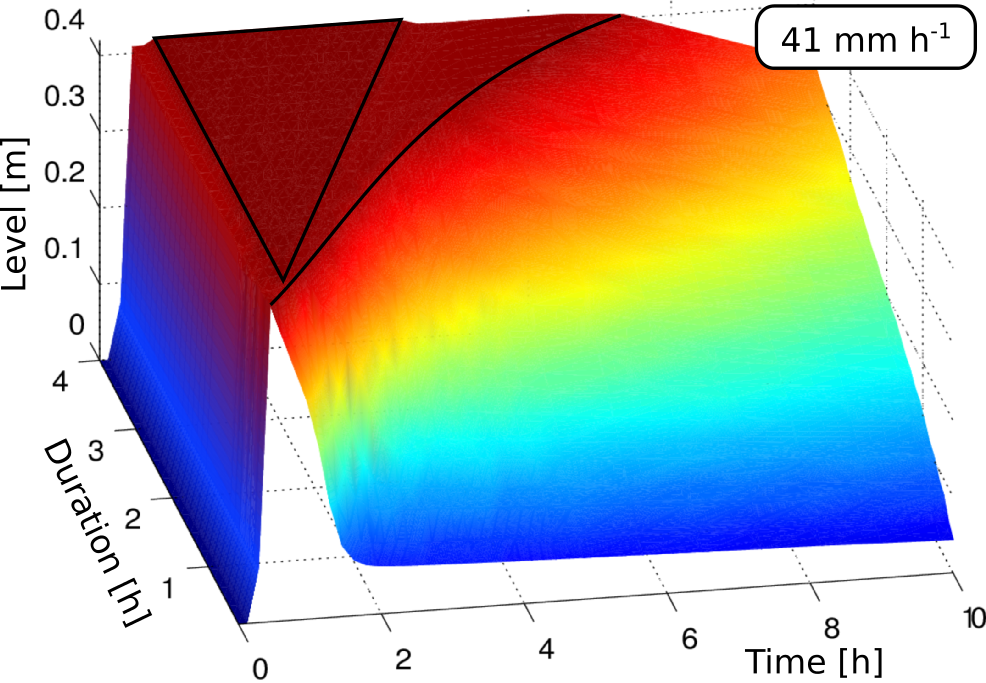}
\protect\caption{Simulator output for a fixed intensity (\SI{41}{\milli\metre\per\hour}) and varying duration.
The black bold lines label different runoff regimes: free-surface flows, runoff with activated storage and emptying of storage in
the catchment.}
\label{fig:wartegg_out}
\end{figure}

Results of these simulations can be seen in Fig.~\ref{fig:wartegg_out}.
For rains shorter than a certain duration (about \SI{20}{\minute} for the intensity used in the plot, \SI{41}{\milli\metre\per\hour}) the water level response shows the typical wave of runoff in an open channel.
After some critical duration, which corresponds to the catchment's time of concentration, the water level becomes constant and remains  fixed for the duration of the rain, defining the triangular region marked in the plot.
After the rain, the water level goes back to a lower fixed level and remains there for a period of time which is a nonlinear function of the rain duration, e.g. between \num{4}-\SI{6}{\hour} for a \SI{4}{\hour} event.
This shows the nonlinear nature of the storage involved in the effective discharge of the network.
Finally there is a slow decay in the level fueled by the residual water in the network, the rate of this decay also depends nonlinearly on the rain's duration.

\begin{figure*}[htbp]
\centering
\includegraphics[width=\columnwidth]{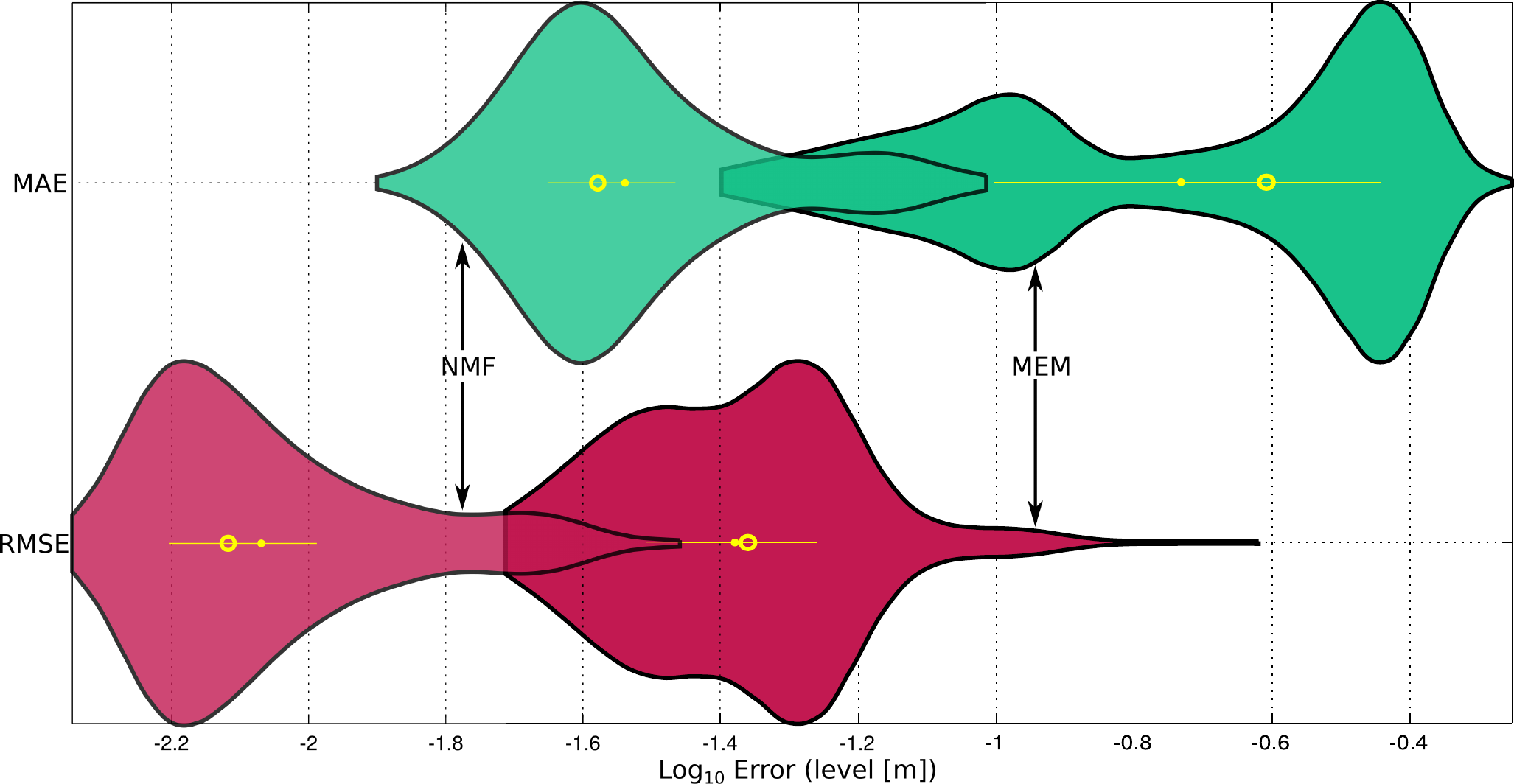}
\protect\caption{Test error distribution of the NMF emulator.
The violin plots show the distribution of the logarithm of the error for the maximum absolute error (MAE) and the root mean square error (RMSE).
Light colored plots corresponds to the NMF emulator, solid colored plots to the MEM.
The mean(median) error is indicated with filled(empty) circles.
See Table~\ref{tab:wartegg_error} for their numerical values.}
\label{fig:wartegg_nmf_error}
\end{figure*}

Fig.~\ref{fig:wartegg_nmf_error} shows the distribution of the test error of a MEM\footnote{Warped MEM, see Sec.~\ref{sec:wartegg_m}. MEMs with linear proxies were unable to reduce average RMSE below 25\%.} and a NMF emulator with $7$ components.
Table~\ref{tab:wartegg_error} summarizes the mean of the error distributions.
The NMF emulator outperforms the mechanistic emulator.
In previous trials, an SVD emulator was also built and provided results comparable with NMF (not shown here), however it produced negative predictions just before the steep increase of the water level at the beginning of the rain event.

\begin{table}[ht]
\begin{tabular}{lcc}
Emulator & MAE (\si{\metre}, \%) & RMSE (\si{\metre}, \%)\\
\hline
NMF(\num{1e4}) & \num{3.2e-2}, 7.7  & \num{0.94e-2}, 4.1\\
MEM(\num{1e2}) & \num{22.6e-2}, 53.7 & \num{4.4e-2}, 17.4
\end{tabular}
\caption{\label{tab:wartegg_error} Mean emulation errors corresponding to the Wartegg catchment dataset.
The value in parenthesis is the simulation speed-up factor obtained with respect to the original simulator, e.g. if SWMM simulation takes \SI{3}{\second}, MNF emulator takes \SI{0.3}{\milli\second}.}
\end{table}

\begin{figure*}[htbp]
\centering
\includegraphics[width=\columnwidth]{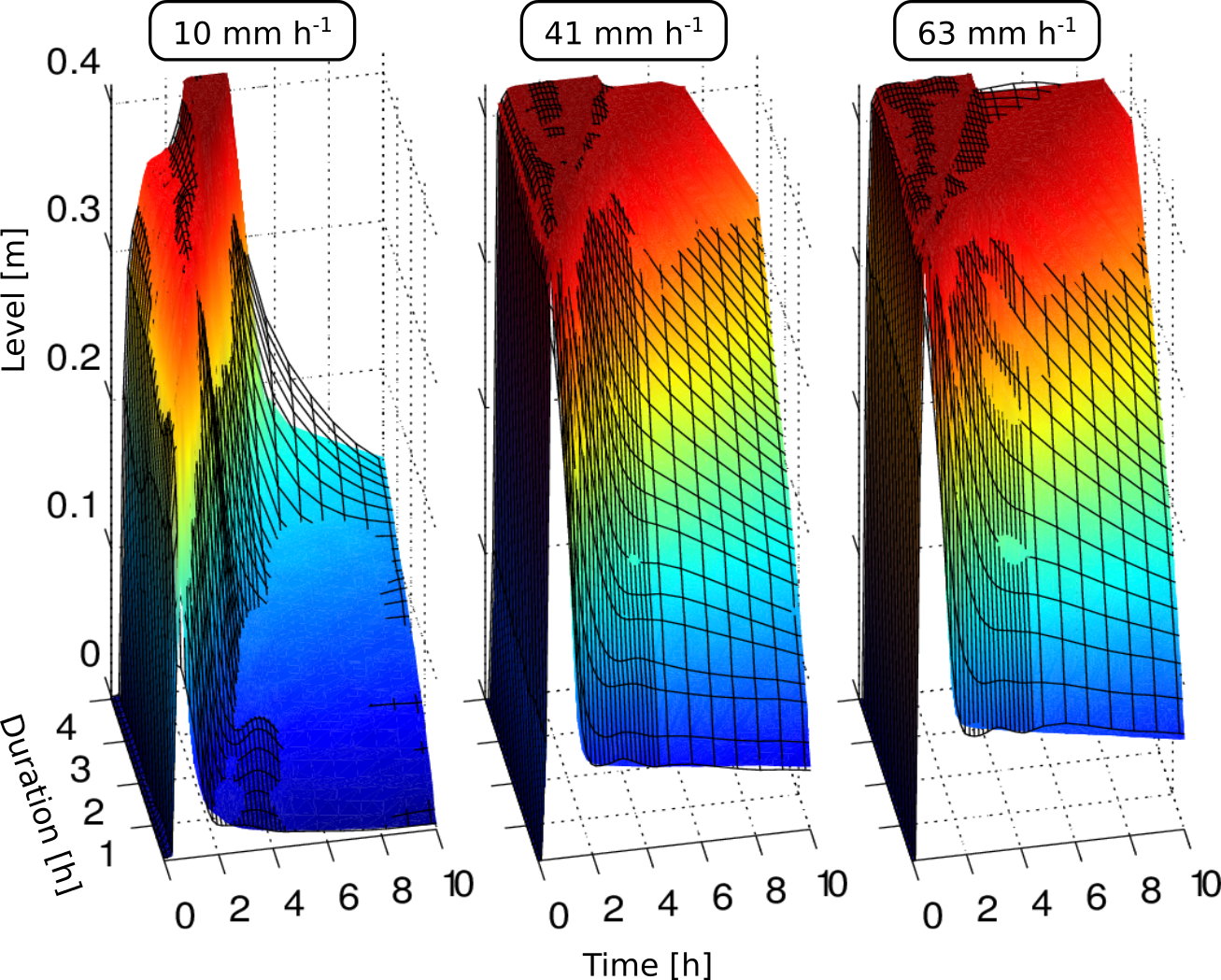}
\protect\caption{NMF emulation quality.
The colored surface shows the simulation outputs for three different rain intensities.
The NMF emulated surface is shown with a black wireframe.}
\label{fig:wartegg_nmf_emu}
\end{figure*}

Fig.~\ref{fig:wartegg_nmf_emu} shows the quality of the NMF emulation for three different rain intensities.

\subsection{Adliswil catchment dataset}
\label{sec:adliswil}
In~\citep{Machac2016} a SWMM model of the Adliswil catchment was emulated using a MEM with a prior derived from a simplified version of the simulator's equations, with the aim of running a system identification task from a single rain event.

\begin{figure}[htbp]
\centering
\includegraphics[width=\columnwidth]{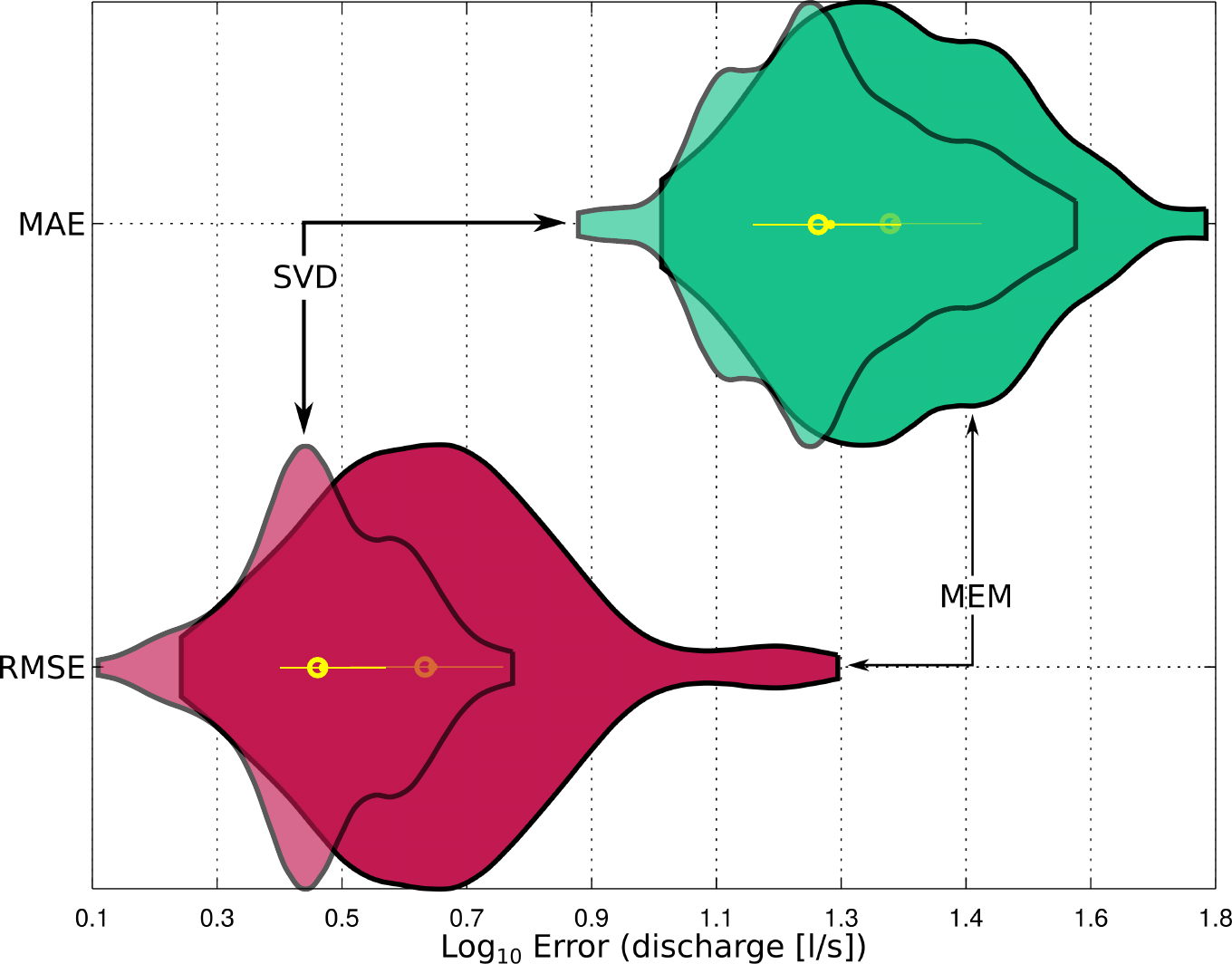}
\protect\caption{Test error distribution of the MEM from~\citet{Machac2016} and a SVD emulator.
The violin plots show the error distribution for the mean maximum absolute error (MAE) and the root mean square error (RMSE).
Light colored plots corresponds to the SVD emulator, solid colored plots to the MEM.
The mean(median) error is indicated with filled(empty) circles.
See Table~\ref{tab:adlis_error} for their numerical values.}
\label{fig:adlis_error}
\end{figure}

Figure~\ref{fig:adlis_error} shows the distribution of the test error of a MEM and a SVD emulator with 6 components.
Both errors are calculated on the same test set.

Table~\ref{tab:adlis_error} summarizes the mean of the error distributions.
In this example, the SVD emulator also performs better than its mechanistic counter part.

Although the time interpolation of SVD does not respect the dynamics of the system, the density of data is enough to provide a good estimation.
The simplicity of the SVD emulator, when compared with the mechanistic one, makes this approach much more compelling for this application.

\begin{table}[ht]
\begin{tabular}{lcc}
Emulator & MAE (\si{\litre\per\second}, \%) & RMSE (\si{\litre\per\second}, \%)\\
\hline
SVD(\num{5e4}) & 18.3, 6.2 & 3.19, 2.6\\
MEM(\num{1e3}) & 23.2, 7.9 & 4.94, 4.0\\
MEM-fit (\num{1e3}) & 20.2, 6.9 & 3.42, 2.8
\end{tabular}
\caption{\label{tab:adlis_error} Mean emulation errors corresponding to the Adliswil catchment dataset.
MEM refers to an emulator in~\citet{Machac2016}, while MEM-fit to an emulator with the same proxy structure but parameters fitted to the data.
The value in parenthesis is the simulation speed-up factor obtained with respect to the original simulator, e.g. if SWMM simulation takes \SI{3}{\second}, SVD emulator takes \SI{60}{\micro\second}.}
\end{table}

\section{Discussion}
\label{sec:discussion}

The results presented in the previous sections seem to suggest that MEMs are not useful for
emulating complicated hydrological or hydrodynamic simulators and that they remain as an academic curiosity.
However MEMs still have many properties that can be exploited for better and faster emulation, which should not be ignored only due to the early state of the method.
These known advantages include the suitability for parallelization and, speedups and energy saving via approximated computing~\citep{Angerer2015}.

Table~\ref{tab:method_comp} summarizes the steps involved in the two approaches presented here.
There we indicate the relation between the method and the characteristics of the dataset.
On one side, although data-driven emulators are computationally and conceptually simpler than MEMs, they are more sensitive to the sparseness of the data.
On the other side, MEM's performance is limited by the linearity of the prior which might fail to express our knowledge about the nonlinear simulator.
If good prior knowledge is available, the mechanistic emulation can incorporate correct dependencies, which could be exploited to reduce the amount of data needed to achieve a desired performance.

Steps 1 and 2 of mechanistic emulation are the most sensitive to the mismatch between prior and simulator.
To mitigate this, we keep the model structure provided by the prior and learn the parameter values from the data, thus providing the best linear proxy for the dataset (Sec.~\ref{sec:staticnlC}).
This generally provides sharper error distributions than using the parameter values obtained directly from the prior~\citep{Albert2012}.
Note that mitigating this simulator-prior mismatch is not a question of data quantity, since more data will override the prior and all mechanistic insights it provided, thus rendering mechanistic knowledge unnecessary~\citep{Steinke2008}.
More data would also increase the memory and computational resources needed by the emulator.
Increasing the density of time samples will also reduce the condition number of the covariance matrices and the inversion problem at the training phase will be ill-posed, unless iterative condition methods are used~\citep{Reichert2011}.

In step 2 of a data-driven emulator we need to factorize the data.
If the data is very sparse the generalization quality of the factorization is expected to be poor.
For data that is sampled at different temporal grids the situation becomes even more delicate since data preprocessing is required to build up a grid, e.g. via interpolation or matrix completion.
Matrix factorization also provide features only at the observed inputs, therefore an interpolation method is required when emulating unseen time points at step 5.
The Kalman smoothing implementation of the mechanistic emulation used in~\citet{Reichert2011} will be similarly affected.
Optimizing this interpolation can be as hard as using a MEM directly (Sec.~\ref{sec:staticnl}).

\begin{table*}[tb]
\centering
\begin{tabular}{l>{\centering\arraybackslash}m{0.4\textwidth}>{\centering\arraybackslash}m{0.4\textwidth}}
\bf{Step} & \bf{Mechanistic emulation} & \bf{Data-driven emulation}\\
\hline
1 & \hl{Pink}{Define prior} & -- \\
\hline
2 & \hl{Pink}{Obtain emulator parameters} & \hl{Pink}{Factorize the data}\\
\hline
3 & Define/Build parameters mapping & Build parameters mapping\\
\hline
4 & Conditioning & -- \\
\hline
5 (Emulation)& Matrix $\cdot$ vector & Matrix $\cdot$ vector + \hl{Pink}{time interpolation}\\
\end{tabular}
\caption{\label{tab:method_comp} Steps involved in the two emulation approaches described in this article. Data-driven approaches are simpler than mechanistic ones. Step 2 and 5 of data-driven approaches suffer if data is sparse or unevenly sampled. Steps 1 and 2 of the mechanistic approach suffer from the lack of expressiveness of linear priors.}
\end{table*}

Equation~\eqref{eq:avg1} shows that the mean function of the prior GP is given by the solution to the noise-free linear ODE obtained by removing the noisy term of the SDE~\eqref{eq:sys1}.
This suggests a simple improvement of the emulator in which the mean function is replaced with a better approximator of the data.
This is the underlying idea behind the work of~\citet{Gonzalez2014}, in which the actuation affecting the mean of the GP is replaced by a signal generated by the nonlinear part of the model applied to a surrogate trajectory.
In that work however, the surrogate trajectory, e.g. built with matrix factorization, does not encode mechanistic knowledge explicitly.
This knowledge could be introduced via the analytical solution of a nonlinear differential equation, such as the nonlinear Bernoulli ODE, or analytical approximations of more general nonlinear differential equations~\citep{Adomian1991}.
These enhancements would only affect the mean function of the predictive GP, improving extrapolation quality and thus allowing for more sparse training sets.
However, the estimation of uncertainties remains limited by the covariance function associated with the linear prior.
This can be improved by including the mean function parameters in the conditioning step~\citep[][sec. 2.7]{Rasmussen06}.

All these observations are derived from more general results on the reducibility and emulation readinness of general simulators and not only valid for the especific simulators we used here. This is specially true for the data-driven approach, see for example Ch. 5 of~\citet{Quarteroni2016}.

\section{Conclusions}
We provided a comparison of mechanistic and data-driven emulation in several examples pertinent to the field of hydrology and urban water management.
In all of these, data-driven emulation outperforms mechanistic emulation.
The current state of MEMs makes them advantageous to fully data-driven emulators, when the training data is sparse and unevenly sampled.
This is the case when many simulation runs with high temporal resolution are prohibitively expensive, or when adaptive stepsize simulators are used.
If the only objective of emulation is to obtain a fast tool to replace a simulator, there seem to be no advantage in using mechanistic knowledge besides the case of sparse and unevenly sampled data mentioned before.
The gain obtained from enhancements of MEMs discussed here, such as Wiener model proxies, Nonlinear mean functions, and hybrid mechanistic/data-driven emulators should be quantified in relation to the test error of an inexpensive data-driven emulator.

\section*{Acknowledgements}
The authors would like to thank Prof. Peter Reichert for his support during the development of this article.
We thank Dr. David Machac for his emulation results used in Figure~\ref{fig:adlis_error}.
We thank Dr. Frank Blumensaat for sharing with us the SWMM model file of the Wartegg catchment resulting from many data collection campaings.
We thank the developers of~\citetalias{Octave, Sage} and~\citetalias{Inkscape} for their excellent software tools, which were used for this article.
We thank the reviewers for their help improving this manuscript.

\paragraph*{Funding} The research leading to these results has received funding from the European Union’s Horizon 2020 research and innovation programme under grant agreement No 641931 (Centaur).

\paragraph*{Author contributions} \textbf{JPC} developed the software, carried out simulations, data analysis. \textbf{JPC} \& \textbf{JR} wrote this manuscript. \textbf{JPL} adapted SWMM model files for the simulations and helped interpreting the results. The work was done under the active supervision of \textbf{CA} \& \textbf{JR}. All authors copy-edited this manuscript.

\sloppy
\bibliographystyle{unsrtnat}
\bibliography{../References}

\end{document}